%% file: hep-ph_0202109v2.tex
\documentclass[11pt]{article}
\usepackage{graphics}
\usepackage{makeidx}
\usepackage{epsf}
\usepackage{here}
\usepackage{feynmp}
\usepackage{mcite}
\usepackage{amsmath}
\usepackage{pstricks}
\usepackage{pst-node}
\usepackage{feynarts}

\def \ni {\noindent}
\def \be {\begin{equation}}
\def \e {\end{equation}}
\def \bea {\begin{eqnarray}}
\def \ea {\end{eqnarray}}
\def \eps {\epsilon}

\def \ga {\gamma}

\def \no {\nonumber}

\def \ps {p\hspace{-0.43em}/}

\newcommand{\To}[2]{\stackrel{#1}{\hbox to #2 pt{\rightarrowfill}}}

\def\wp{\ifmmode W^+\else $W^+$\fi}
\def\wm{\ifmmode W^-\else $W^-$\fi}
\def\emm{\ifmmode e^-\else $e^-$\fi}
\def\ep{\ifmmode e^+\else $e^+$\fi}

\def\sw{\ensuremath{ \sin \theta_{\rm w}}}
\def\swto{\ensuremath{ \sin^2 \theta_{\rm w}}}

\def\cwto{\ensuremath{ \cos^2 \theta_{\rm w}}}

\def\mw{\ensuremath{ {M}_{\scriptscriptstyle W}} }
\def\mz{\ensuremath{ {M}_{\scriptscriptstyle Z}} }
\def\mh{\ensuremath{ {M}_{\scriptscriptstyle H}} }

\def \mt{\ensuremath{ m_t}}

\def \uo{\ensuremath { u(p_2) \,\, }}
\def \ut{\ensuremath { \bar{u}(p_3) \,\, }}
\def \vo{\ensuremath { v(p_4) \,\, }}
\def \vt{\ensuremath { \bar{v}(p_1) \,\, }}

\def \e#1#2{\ensuremath { \eps_{\,#1}^{\,*}\,({#2})\,\, }}

\def \eett {\ensuremath {e^+e^- \to t \bar{t} \,\,}}
\def \GeV {\ensuremath  \,{\rm GeV} \,\,}

\def\sig{\left[\frac{\displaystyle{\mathrm{d}\sigma}}{\displaystyle{\mathrm{d}\cos \, \theta}}\right]}

\def\unity{{\rm 1\mskip-4.25mu l}}
\def\re{\mathop{\mathrm{Re}}}

\begin{document}

\baselineskip 5mm



\input titlepage_v2

\newpage

\input intro_v2

\input twoprojects_v2

\input numerics_v2

\input conclusion_v2

\newpage

\begin{appendix}
  \input app_v2
\end{appendix}

\newpage

\bibliographystyle{/afs/ifh.de/user/r/riemann/digedag/tex/bib/utphys_}
\bibliography{/afs/ifh.de/user/r/riemann/digedag/tex/bib/toppair}


\end{document}

%% file: titlepage_v2.tex
\vspace{-2cm}
\renewcommand{\thefootnote}{\fnsymbol{footnote}}  
\begin{flushright}  
LC--TH--2002--002\\  
hep-ph/0202109\\
September 2002 \\  
\end{flushright}  
\vspace{1cm} 
\begin{center}  
{\Large \bf Complete  electroweak one-loop radiative corrections 
\\[3mm]
to 
top-pair
  production at TESLA 
\\[3mm]
-- a comparison\footnote{
Work supported in part by the European Community's Human Potential
Programme under contract HPRN-CT-2000-00149 Physics at Colliders.
\\[-1mm]} -- }
  \\ 
\vspace{1.5cm} 
{ 
{\large J. Fleischer}${}^{1}$\footnote{E-mails:~fl@physik.uni-bielefeld.de,
hahn@feynarts.de,
hollik@particle.uni-karlsruhe.de,
Tord.Riemann@desy.de,
cs@particle.uni-karlsruhe.de,
Anja.Werthenbach@desy.de
}}, 
~~{\large T. Hahn}${}^{2}$,
~~{\large W. Hollik}${}^{3,4}$,
~~{\large T. Riemann}${}^{5}$,\\
~~{\large C. Schappacher}${}^{3}$
~~ and
~~{\large A. Werthenbach}${}^{5}$
 \\
\vspace{1cm} 

{\small 
{
${}^{1}$~Universit\"at Bielefeld, Universit\"atsstrasse 25,  D--33615
Bielefeld, Germany\\ }
\smallskip
{
${}^{2}$~Fraunhofer-Institut f\"ur Integrierte Schaltungen,
Am Weichselgarten 3,
D--91058 Erlangen, Germany
\\ }
\smallskip
{
${}^{3}$~Universit\"at Karlsruhe,
    Inst. f\"ur Theoretische Physik,
    Kaiserstrasse 12,
    D--76128 Karlsruhe, Germany\\ }
\smallskip
{
${}^{4}$~Max-Planck-Institut f\"ur Physik,
    F\"ohringer Ring 6,
    D--80805 M\"unchen, Germany\\}
\smallskip
{
${}^{5}$~Deutsches Elektronen-Synchrotron, DESY Zeuthen, Platanenallee
  6, D--15738 Zeuthen, Germany  }  

}
\end{center}

\vspace{1cm} 

\begin{center}
{\bf \sc{Abstract}}
\end{center}
Electroweak one-loop radiative corrections to the process \eett are revisited.
Two groups from Karlsruhe and Bielefeld/Zeuthen performed independent
calculations of both (virtual and soft) QED
contributions and weak virtual corrections. 
For the angular distribution an agreement of at least eight
digits for the weak
corrections and of at least seven digits for additional  photonic
corrections is established.  
\\

\setcounter{footnote}{0}  
\renewcommand{\thefootnote}{\arabic{footnote}}  
\setcounter{page}{1}  
\thispagestyle{empty} 


%% file: intro_v2.tex
\section{Introduction}
At TESLA, one of the most important production processes will be
top-pair production well above the production threshold (in the
continuum)
\begin{eqnarray}
\label{eq1}
  \label{eq:qqtt}
  e^+~(p_1) ~+~ e^-~(p_2) \rightarrow t~(p_3) ~+~  \bar t~(p_4) .
\end{eqnarray}
We expect several hundred thousand events, and the anticipated
accuracy of the theoretical predictions to be a few per mille.
What one has to calculate with such net precision is, of course, not only
the two-fermion production process, but also the decay of the top
quarks and the variety of quite different radiative corrections like
real photonic bremsstrahlung, 
electroweak radiative corrections (EWRC),    
QCD corrections to the final state, and beamstrahlung.
Potentially, new physics effects also have to be taken into
account.\\[2mm]
\ni
The state of the art at the time of the public presentation of the TESLA
project has been reviewed in Part III of the Technical Design Report
 \cite{Aguilar-Saavedra:2001rg}.
Since then, the community has made progress in several directions. 
In this note we report on recent developments in the
description of the electroweak radiative corrections to (\ref{eq1}). 
Several studies on this topic are available in the literature, e.g.  
\cite{Fujimoto:1988hu,Beenakker:1991ca,Hollik:1998md,Bardin:2000kn}, 
but no special effort was undertaken to perform a detailed numerical
comparison between the results of different collaborations.
Quite recently, a recalculation of the electroweak corrections has
been performed by a Bielefeld/Zeuthen group \cite{FRW:2002sw091}, and
this calculation was compared in detail with the Karlsruhe 
results.\footnote{A series of numerical comparisons of the
 Bielefeld/Zeuthen group with the authors of \cite{Bardin:2000kn}
  started in September 2001 and was later joined by the Karlsruhe
 group.} \\[2mm]
\ni
We summarize this comparison together with a few remarks on
the underlying projects.   


%% file: twoprojects_v2.tex
\section{One-loop EWRC to the reaction \boldmath {\eett}}
\newcommand{\bqa}{\begin{eqnarray}}
\newcommand{\eqa}{\end{eqnarray}}
\def\nl{\nonumber \\}
\newcommand{\ba}{\begin{eqnarray}}
\newcommand{\nll}{\nonumber\\}
\newcommand{\nnn}{\\ \nonumber \\}

\ni
In this section we introduce the one-loop differential cross-section of the process \eett in the Standard Model.\\[2mm]
\ni
To start with, we first present the differential Born cross-section, 
\bqa
\frac{\rm{d}\sigma^{\rm{Born}}}{\rm{d}\cos\theta}
&=& 
\left( 1+\cos^2\theta \right) 
\sigma_{{T}}^0(s)
+
2\cos\theta~ 
\sigma_{{FB}}^0(s)
+ \frac{4\,m_t^2}{s}\,
\left(1-\cos^2\theta\right)~ 
\sigma_{{T}}^{0,m}(s) ,
\label{sigdiff}
\eqa

\ni
with

\begin{align}
\sigma_{{T}}^0(s) &=
\frac{\pi \alpha^2}{2s} c_t \beta_t
\left[ 
Q_e^2 Q_t^2         
+
2Q_eQ_tv_ev_t 
\,\re\chi_{{Z}}(s)
+
\bigl(v_e^2+a_e^2\bigr)
                    \bigl(v_t^2+a_t^2\beta_t^2\bigr) 
\,|\chi_{{Z}}(s)|^2 \right],
\\ \nl 
\sigma_{{FB}}^0(s) 
&=
\frac{\pi \alpha^2}{2s} c_t \beta_t^2 
\left[ 
2Q_eQ_t a_ea_t 
\re \chi_{{Z}}(s)
+
4v_ea_ev_ta_t 
\left|\chi_{{Z}}(s)\right|^2\right]\,,
\\ \nl 
\sigma_{{T}}^{0,m}(s) &=  
\frac{\pi \alpha^2}{2s} c_t \beta_t
\left[ 
Q_e^2Q_t^2  
      +
2Q_eQ_tv_ev_t 
\,\re\chi_{{Z}}(s)
      +
\bigl(v_e^2+a_e^2\bigr)v_t^2 
\, |\chi_{{Z}}(s)|^2 \right]\,.
\label{sigdiff1}
\end{align}
\ni
The $Z$ propagator is contained in the factor
\ba
\chi_{{Z}}(s) &=& 
\frac{g^2}{4 \, e^2 \cwto }  
~~\frac{s}{s-m^2_{{Z}}},
\label{chiZ}
\\ \nl
m^2_{{Z}} &=& \mz^2 - i \mz \Gamma_Z,
\label{mZ2}
\ea

\ni
and we use the following conventions
\bqa
\label{kg1}
Q_e&=&-1, 
\\
\label{kg2}
a_f&=&I_3^L(f)~~=~~\pm \frac{1}{2}, 
\\
\label{kg3}
v_f&=& a_f(1 - 4 |Q_f| \swto),
\eqa
together with
\bqa
\beta_t &=& \sqrt{1-\frac{4\,m_t^2}{s}}\,,
\label{mus}
\eqa
\ni and the colour factor $c_t=3$ of the top quark.  
\ni
Including the virtual electroweak corrections and the soft photonic contributions 
 it proves to be convenient to use the notation of complex-valued form 
factors, each of them related to some helicity-based matrix element.
For the production of massive fermion pairs, 
the cross-section depends on six independent form factors,
\begin{align}
\frac{{\rm d} \sigma}{{\rm d} \cos \, \theta} & =  \,\frac{\beta_t}{16\pi \, s} \,  | {\cal M}_{\gamma + Z } |^2,
\end{align}
with


\begin{align}
\label{Xsec}
| {\cal M}_{\gamma + Z } |^2 = &  
~\left( U^2 + T^2 \right) 
\left( |{\rm F_1^{11}}|^2 + |{\rm F_1^{15}}|^2 + |{\rm F_1^{51}}|^2
 + |{\rm F_1^{55}}|^2   \right)       
\\ \no
 & +~ 2 \left( U^2 - T^2 \right) 
\re 
\left({\rm F_1^{11}}{\rm F_1^{55}}^{*} + {\rm F_1^{15}}{\rm F_1^{51}}^{*} 
\right)
\\ \no
 & + ~2 \, \mt^2 \,s \left( |{\rm F_1^{11}}|^2   - |{\rm F_1^{15}}|^2
+  |{\rm F_1^{51}}|^2  -  |{\rm F_1^{55}}|^2 \right)
\\ \no
  & +~2\,\frac{m_t^2}{s}(U\,T-m_t^4) 
\re  \left(  {\rm F_3^{11}}  {\rm F_1^{11}}^* 
+  {\rm F_3^{51}}  {\rm F_1^{51}}^* \right) \,+  {\cal O}(\alpha^2)\,,
\end{align}
with $s=(p_1+p_2)^2$ and 
\begin{align}
T & = \frac{s}{2} \, (1-\beta_t\, \cos \, \theta)  ,
\\
U & =  \frac{s}{2} \, (1+\beta_t \, \cos \, \theta) .
\end{align}

\ni
The form factors appear in the decomposition of the 
 matrix element according to
\begin{align}
 \, {\cal M}_{\gamma + Z } & = 
\sum_{a,b=1,5} {\rm F_1}^{ab} \,\,    {\cal M}_{1,{ab}} +  
 {\rm F_3^{11}} \,\,    {\cal M}_{3,{11}} +  
 {\rm F_3^{51}} \,\,    {\cal M}_{3,{51}}.
\end{align}
The matrix elements ${\cal M}_{3,{11}}$ and ${\cal
M}_{3,{51}}$ emerge only in the calculation of the virtual corrections.
 Their interference with
the Born amplitudes vanishes
in the case of massless fermion pair production.
The other matrix elements $ {\cal M}_{1,{ab}} $ receive additional contributions from soft photon 
corrections.  
In Born approximation, the non-vanishing form factors are
\begin{align}
 {\rm F_{1,B}^{11}} & = \left(e^2\, \chi_{{Z}}(s) \, v_e \, v_t   + e^2 \, Q_e
 \, Q_t \right) \, / \, s, 
\\
 {\rm F_{1,B}^{15}} & = - e^2\,\chi_{{Z}}(s) \, v_e \, a_t \, / \, s , 
\\
 {\rm F_{1,B}^{51}} & = - e^2\,\chi_{{Z}}(s) \, v_t \, a_e  \, / \, s,
\\
 {\rm F_{1,B}^{55}} & =   e^2\,\chi_{{Z}}(s) \, a_e \, a_t  \, / \, s.
\end{align}
\ni
For completeness, we give also the matrix elements
\begin{align}
i \, {\cal M}_{1\,{ab}} & = [\vt \ga^{\mu}\, g_a \, \uo] \,\, [\ut
\ga_{\mu}\, g_b \, \vo] ,~~~~a,b=1,5,
\end{align}
where $g_1=1$ and $g_5=\gamma_5$.
For massive loop corrections, two additional structures have to be added
\begin{align}
i \, {\cal M}_{3,{11}} & = [\vt  \unity \, \uo] \,\, [\ut \ps_2\, \unity \, \vo]  ,
\\
i \, {\cal M}_{3,{51}} & = [\vt \ps_3\, \ga_5 \, \uo] \,\, [\ut \unity \, \vo]. 
\end{align}

%
%
%

\ni
In Appendix A we give the collection of all Feynman
diagrams relevant at one-loop level for the calculation in the 't Hooft--Feynman
gauge.
Ultra-violet (UV) divergences were treated within dimensional
   regularization. For renormalization the on-shell scheme
 was used \cite{Bardin:1997xq}.
\ni
Infra-red (IR) divergences cancel analytically between the soft
photon emission and the diagrams with virtual photons. 
This leads to a residual dependence of the cross-section on the maximal 
soft-photon energy, which has been arbitrarily chosen to be 
$E_{\gamma}^{\max} = \sqrt{s}/10$. 
Of course, after including real hard photonic bremsstrahlung this dependence 
disappears.\\[2mm]
\ni
In the on-shell renormalization scheme the weak 
mixing angle is defined by the weak boson masses. 
In this calculation the coupling constant  $g$ (\ref{chiZ}) is
expressed by the electric charge and the weak mixing angle.
\begin{eqnarray}
\label{eq2}
\sin^2\theta_W &=& 1-\frac{\mw^2}{\mz^2},
\\
\label{eq3}
g &=& \frac{e}{\sw} \,.
\label{eq:qqqqqq}
\end{eqnarray}
\label{eq:qqw}


\subsection{The Karlsruhe Approach}






   The Karlsruhe calculations have a long history dating back to the
   late 1980s, following the formulation worked out in \cite{Bohm:1986rj,Denner:1993kt}. We shall not describe them here in detail. They use
the
   software packages {\tt FeynArts} \cite{Kublbeck:1990xc,Hahn:2000kx}, {\tt FormCalc} \cite{Hahn:1998yk}, and
   {\tt LoopTools} \cite{Hahn:1998yk}, the latter of which is based on the
{\tt FF}
library
   \cite{vanOldenborgh:1991yc}. The calculations follow the scheme described in
   \cite{Hahn:2000jm}: the diagrams are first generated with FeynArts, the
   resulting amplitudes are analytically simplified with
FormCalc,
   whose output is then converted to a Fortran program which is
linked
   with the LoopTools library to produce e.g. cross-sections. This
   process takes only a few minutes and is highly automated.\\[2mm]
\ni
   Results of calculations performed in this way have been made
   available on {\it The HEP Process} {\it  Repository} Web page \cite{HEP-repository}. The
   actual codes used for comparison in this paper have been taken
from
   there. In particular we used the files \verb=eett_sm.tar.gz= and
   \verb=eett_smnoqed.tar.gz=.\\[2mm]
\ni
We just mention that a file with the virtual one-loop corrections in
the MSSM,  
{\tt eett\_mssm.tar.gz},  is also obtainable.


\subsection{The Bielefeld/Zeuthen Approach}

The Bielefeld/Zeuthen group utilized the program package {\tt DIANA} 
\cite{Tentyukov:1999is,Fleischer:2000zr}, which is a {\tt FORM} interface
for the package {\tt qgraph} \cite{Nogueira:1993ex} 
for the creation of all the
contributing Feynman diagrams to a given process.\\[2mm]
\ni
The output of {\tt DIANA} is a sample of symbolic expressions which then
were prepared for the integration of the loop momenta with 
{\tt FORM} \cite{FORM}.
 Two independent calculations which differ in many respects have been performed.
In one of them,  dimensional
regularization is used for UV and IR divergences, while in the other 
 the IR divergences are regularized by a finite, small photon mass. 
The resulting expressions in terms of Passarino-Veltman functions are
calculated with a Fortran code, using two libraries of one-loop functions:
the dimensionally-regularized version uses {\tt LoopTools} \cite{Hahn:1998yk} with 
minor additions, the other one the package {\tt FF} \cite{vanOldenborgh:1991yc}. 
%
%
%
On-shell renormalization is performed following \cite{Fleischer:1981ub}.
To eliminate the IR divergence of the virtual photonic
contributions  soft photon bremsstrahlung has been added; 
 we followed the techniques described in \cite{Akhundov:1996my}.\\[2mm]
\ni
The Fortran program {\tt topfit} provides two possible outputs:
Firstly the differential cross-section (and derived observables like
total cross-section or forward-backward asymmetry), secondly six independent 
form factors in a given helicity basis, as introduced above,  
designed to be used in a Monte-Carlo program. \\[2mm]
\ni
It is worth mentioning that {\tt topfit} is equipped to calculate the 
hard real photon corrections to the differential cross-section.


%% file: numerics_v2.tex
\section{Numerical Results \label{comparison}}
The comparison of the two calculations is chosen to be performed
on the level of the differential cross-section. 
We focus in particular on the following four contributions
\begin{itemize}
 \item $\sig_{\rm Born}$ : Born cross-section which serves as a cross
   check of input parameters and conventions.
  \item $\sig_{\rm QED}$ : These numbers contain the Born cross-section plus the interference of Born 
(s-channel
  $\gamma$  and Z exchange) with one-loop virtual QED diagrams plus
the absolute square of real soft photon radiation.
   No vacuum polarization
   diagrams, nor counter--diagrams, of the photon are taken into account, i.e. the running of the electromagnetic
   coupling constant is not included.
   It is assumed that this
   can be easily accounted for, including the variety of higher order 
   corrections, by using the corresponding value of
   $\alpha$. 
 \item $\sig_{\rm weak}$ :  The interference of Born (s-channel
   $\gamma$ and Z exchange) with one-loop virtual pure weak diagrams
is shown.
 As before, the Born cross-section is added to the one-loop correction. The running of the electromagnetic
   coupling constant is included here.
 Renormalization is performed in the
 on-shell scheme and self-energy diagrams are taken into account as well. 

\item $\sig_{\rm SM}$ : All previous parts of the calculation are put
  together, i.e. the complete electroweak one-loop plus Born differential
  cross-section is given within the Standard Model.\\
\end{itemize}

\ni
All numbers in the following tables were obtained making use of the
input parameters  
\begin{align}
\begin{array}[r]{lll}
  \Gamma_Z = 0   \, \,,  & \alpha = \frac{e^2}{4\, \pi}
= 1 / 137.03599976 \,, & E_{\gamma}^{\max} = \sqrt{s}/10 \,, \\[2mm]
 \mw = 80.4514958 \GeV \, , & \mz = 91.1867 \GeV \,, & \mh = 120 \GeV \,,
 \\[1mm]
 m_e= 0.00051099907 \GeV \,, & m_t = 173.8 \GeV \,, & m_b = 4.7 \GeV \,,
 \\[1mm]
 m_{\mu} = 0.105658389 \GeV \,, & m_u = 0.062 \GeV \,, & m_d = 0.083
\GeV  \,, \\[1mm]
  m_{\tau} = 1.77705 \GeV \,, & m_c = 1.5 \GeV \,, & m_s = 0.215 \GeV \,.
\end{array}
 \end{align}
\ni
Effective quark masses reproducing the hadronic vacuum polarization
contribution $\Delta \alpha_h$ with a sufficiently high accurancy
have been chosen \cite{Eidelman:1995ny,Bardin:1997xq}. 
\ni
As mentioned, the photon energy cut $E_{\gamma}^{\max}$ is an arbitary
quantity entering the final result if the radiation of hard photons in
not taken into account.\\[2mm]
\ni
The numerical values of the differential cross-sections have been
calculated at three typical Next--Linear--Collider energies.\\[-8mm] 
\input tables_v2   
\ni
From Table 1 it is evident that the numerics of the weak virtual corrections are perfectly 
controlled with a gross agreement of the two approaches by at least eight 
digits.\\[2mm]
\ni
The pure photonic corrections agree to at least seven digits%
\footnote{The {\tt topfit} part was analytically and numerically
cross-checked against \cite{Beenakker:thesis}.}.
The net result of the comparison being highly satisfactory, we did not
push for more digits agreement. \\[2mm]
\ni
Finally, it has to be pointed out that we control here only what is often called the 
{\em technical precision} of a calculation, not to be confused with the precision 
of the prediction of some observable to be confronted with some realistic 
measurement.




%% file: tables_v2.tex



\vspace*{5mm}

$\sqrt s = 500$ GeV:
$$
\begin{array}{|l|l|l|l|l|}
\hline 
\vrule height 3ex depth 0ex width 0ex
\cos\theta & \sig_{\text{Born}} & \sig_{\text{QED}} & 
\sig_{\text{weak}} & \sig_{\text{SM}} \\ \hline 
\hline 
-0.9 &  0.1088~3919~4075   &  0.09866~4252         &  0.1242~59037~1      & 0.1140~8410
\\
     &  0.1088~3919~4075   &  0.09866~4253         &  0.1242~59037~6      & 0.1140~8410
\\\hline 
-0.5 &  0.1422~7506~9392   &  0.1285~0790          &  0.1568~48371~9      & 0.1430~8121 
\\
     &  0.1422~7506~9392   &  0.1285~0790          &  0.1568~48371~8      & 0.1430~8121
\\\hline 
0.   &  0.2254~7046~4032   &  0.2023~9167          &  0.2402~6680~4       & 0.2171~8801 
\\
     &  0.2254~7046~4033   &  0.2023~9167          &  0.2402~6680~3       & 0.2171~8801
\\\hline 
0.5  &  0.3546~6647~0332   &  0.3151~1724          &  0.3688~8650~7       & 0.3293~3727 
\\
     &  0.3546~6647~0332   &  0.3151~1723          &  0.3688~8650~5       & 0.3293~3727
\\\hline 
0.9  &  0.4911~4371~5766   &  0.4307~1437          &  0.5033~3751~2       & 0.4429~0817
\\
     &  0.4911~4371~5767   &  0.4307~1437          &  0.5033~3750~8       & 0.4429~0816
\\ \hline
\end{array}
$$
\bigskip

$\sqrt s = 700$ GeV:
$$
\begin{array}{|l|l|l|l|l|}
\hline 
\vrule height 3ex depth 0ex width 0ex
\cos\theta & \sig_{\text{Born}} & \sig_{\text{QED}} & 
\sig_{\text{weak}} & \sig_{\text{SM}} \\ \hline \hline 
-0.9 &  0.05033~2867~3357 &  0.04534~6950        &  0.05697~6919~7    & 0.05199~1003        
\\
     &  0.05033~2867~3357    &  0.04534~6948        &  0.05697~6920~1    & 0.05199~1001
\\ \hline
-0.5 &  0.06658~1166~2000    &  0.05993~8980        &  0.07224~8491~7    & 0.06560~6305
\\
     &  0.06658~1166~2001    &  0.05993~8980        &  0.07224~8491~7    & 0.06560~6305
\\ \hline
0.   &  0.1237~4052~8515     &  0.1106~4932         &  0.1280~5635~7     & 0.1149~6514 
\\
     &  0.1237~4052~8515     &  0.1106~4932         &  0.1280~5635~6     & 0.1149~6514
\\ \hline
0.5  &  0.2218~4321~1646     &  0.1955~2864         &  0.2224~6611~4     & 0.1961~5154 
\\
     &  0.2218~4321~1646     &  0.1955~2864         &  0.2224~6611~3     & 0.1961~5154
\\ \hline
0.9  &  0.3298~0454~9138     &  0.2844~3984  &  0.3245~1514~8     & 0.2791~5044        
\\ 
     &  0.3298~0454~9138     &  0.2844~3985     &  0.3245~1514~5     & 0.2791~5045    
\\ \hline
\end{array}
$$
\bigskip
\begin{table}[H]
$\sqrt s = 1000$ GeV:
$$
\begin{array}{|l|l|l|l|l|}
\hline 
\vrule height 3ex depth 0ex width 0ex
\cos\theta & \sig_{\text{Born}} & \sig_{\text{QED}} & 
\sig_{\text{weak}} & \sig_{\text{SM}} \\ \hline \hline 
-0.9 &  0.02278~5423~2732   &  0.02036~5843    &  0.02552~1285 & 0.02310~1705 
\\
     &  0.02278~5423~2732   &  0.02036~5844    &  0.02552~1286   & 0.02310~1706
\\ \hline
-0.5 &  0.02978~2131~1031   &  0.02674~1661    &  0.03186~3489& 0.02882~3019
\\
     &  0.02978~2131~1031   &  0.02674~1663    &  0.03186~3490   & 0.02882~3021
\\ \hline
0.   &  0.06118~0067~4224   &  0.05453~9344    &  0.06159~1613& 0.05495~0889
\\
     &  0.06118~0067~4224   &  0.05453~9344    &  0.06159~1613   & 0.05495~0889
\\ \hline
0.5  &  0.1177~4694~9888    &  0.1031~1627     &  0.1140~47686 & 0.09941~7009 
\\
     &  0.1177~4694~9888    &  0.1031~1626     &  0.1140~47685    & 0.09941~6999 
\\ \hline
0.9  &  0.1811~2209~7086    &  0.1540~3824     &  0.1713~46193 & 0.1442~6233 
\\
     &  0.1811~2209~7086    &  0.1540~3823     &  0.1713~46191    & 0.1442~6232
\\ \hline
\end{array}
$$

\caption{The content of the tables is explained in the text, the upper and lower numbers correspond to the
  Karlsruhe and Bielefeld/Zeuthen approach, respectively.
}
\end{table}

%% file: conclusion_v2.tex
\section{Summary}

We have demonstrated an agreement of seven to eight digits between the 
calculations of the Karlsruhe group and the Bielefeld/Zeuthen group
for electroweak virtual and soft photonic corrections to the reaction \eett 
at the one-loop level.\\[2mm]
\ni
Naturally the claim of completeness of this calculation can only be
applied to the {\it  electroweak} corrections and only at
one-loop accuracy. 
In particular the inclusion of QCD corrections \cite{Aguilar-Saavedra:2001rg} and the finite life time of 
the top quarks have to be taken into account.
To what extent higher-order weak corrections will be needed depends 
crucially on the expected experimental accuracy.
Certainly, it is desirable to partially extend the
calculations to higher-order effects, where a special emphasis could be
given to large mass effects \cite{Bardin:1997xq} and to the so-called Sudakov logarithms \cite{Beenakker:2001kf}, which 
constitute the leading
contributions at the next order in perturbation theory. 
Finally, 
needless to mention that various higher-order photonic
corrections have to be added in high-precision numerical approach.

\section{Acknowledgments}
We would like to thank D. Bardin, P. Christova, and L. Kalinovskaya for
informal discussions.


%% file: app_v2.tex
\section*{Appendix}
\label{app}
\addcontentsline{toc}{section}{Appendix}
\def\thesection{\Alph{section}}
\def\theequation{\thesubsection.\arabic{equation}}
\setcounter{subsection}{0}

\section{Feynman diagrams}

In cases where several particles are listed for a propagator,
there is one diagram for each combination of particles. On fermion
lines, an $f_i$ stands for $\{e, \mu, \tau, u, c, t, d, s, b\}$ 
and a $\nu_i$ for $\{\nu_e, \nu_\mu, \nu_\tau\}$.\\

\ni
Born diagrams:\\[-8mm]
   \input{born}

\ni
Counter-term diagrams:\\[-8mm]
   \input{counter}

\ni
Self-energy diagrams:\\[-8mm]
   \input{self}

\newpage
\ni
Vertex diagrams:\\[-8mm]
   \input{vert}

\ni
Real-photon-emission diagrams:\\[-8mm]
   \input{real}

\ni
Box diagrams:\\[-8mm]
   \input{box}


%% file: born.tex
\unitlength=1bp%

\begin{scriptsize}
\begin{feynartspicture}(432,101)(4,1)
\FADiagram{}
\FAProp(0.,15.)(6.,10.)(0.,){/Straight}{-1}
\FALabel(2.48771,11.7893)[tr]{$e$}
\FAProp(0.,5.)(6.,10.)(0.,){/Straight}{1}
\FALabel(3.51229,6.78926)[tl]{$e$}
\FAProp(20.,15.)(14.,10.)(0.,){/Straight}{1}
\FALabel(16.4877,13.2107)[br]{$t$}
\FAProp(20.,5.)(14.,10.)(0.,){/Straight}{-1}
\FALabel(17.5123,8.21074)[bl]{$t$}
\FAProp(6.,10.)(14.,10.)(0.,){/Sine}{0}
\FALabel(10.,8.93)[t]{$\gamma, Z$}
\FAVert(6.,10.){0}
\FAVert(14.,10.){0}
\end{feynartspicture}
\end{scriptsize}

%% file: counter.tex
\unitlength=1bp%

\begin{scriptsize}
\begin{feynartspicture}(432,101)(4,1)
\FADiagram{}
\FAProp(0.,15.)(6.,10.)(0.,){/Straight}{-1}
\FALabel(2.48771,11.7893)[tr]{$e$}
\FAProp(0.,5.)(6.,10.)(0.,){/Straight}{1}
\FALabel(3.51229,6.78926)[tl]{$e$}
\FAProp(20.,15.)(14.,10.)(0.,){/Straight}{1}
\FALabel(16.4877,13.2107)[br]{$t$}
\FAProp(20.,5.)(14.,10.)(0.,){/Straight}{-1}
\FALabel(17.5123,8.21074)[bl]{$t$}
\FAProp(10.,10.)(6.,10.)(0.,){/Sine}{0}
\FALabel(8.,8.93)[t]{$\gamma, Z$}
\FAProp(10.,10.)(14.,10.)(0.,){/Sine}{0}
\FALabel(12.,11.07)[b]{$\gamma, Z$}
\FAVert(6.,10.){0}
\FAVert(14.,10.){0}
\FAVert(10.,10.){1}

\FADiagram{}
\FAProp(0.,15.)(6.,10.)(0.,){/Straight}{-1}
\FALabel(2.48771,11.7893)[tr]{$e$}
\FAProp(0.,5.)(6.,10.)(0.,){/Straight}{1}
\FALabel(3.51229,6.78926)[tl]{$e$}
\FAProp(20.,15.)(14.,10.)(0.,){/Straight}{1}
\FALabel(16.4877,13.2107)[br]{$t$}
\FAProp(20.,5.)(14.,10.)(0.,){/Straight}{-1}
\FALabel(17.5123,8.21074)[bl]{$t$}
\FAProp(6.,10.)(14.,10.)(0.,){/Sine}{0}
\FALabel(10.,8.93)[t]{$\gamma, Z$}
\FAVert(6.,10.){0}
\FAVert(14.,10.){1}

\FADiagram{}
\FAProp(0.,15.)(6.,10.)(0.,){/Straight}{-1}
\FALabel(2.48771,11.7893)[tr]{$e$}
\FAProp(0.,5.)(6.,10.)(0.,){/Straight}{1}
\FALabel(3.51229,6.78926)[tl]{$e$}
\FAProp(20.,15.)(14.,10.)(0.,){/Straight}{1}
\FALabel(16.4877,13.2107)[br]{$t$}
\FAProp(20.,5.)(14.,10.)(0.,){/Straight}{-1}
\FALabel(17.5123,8.21074)[bl]{$t$}
\FAProp(6.,10.)(14.,10.)(0.,){/Sine}{0}
\FALabel(10.,8.93)[t]{$\gamma, Z$}
\FAVert(14.,10.){0}
\FAVert(6.,10.){1}
\end{feynartspicture}
\end{scriptsize}

%% file: self.tex
\unitlength=1bp%

\begin{scriptsize}
\begin{feynartspicture}(432,303)(4,3)
\FADiagram{}
\FAProp(0.,15.)(4.,10.)(0.,){/Straight}{-1}
\FALabel(1.26965,12.0117)[tr]{$e$}
\FAProp(0.,5.)(4.,10.)(0.,){/Straight}{1}
\FALabel(2.73035,7.01172)[tl]{$e$}
\FAProp(20.,15.)(16.,10.)(0.,){/Straight}{1}
\FALabel(17.2697,12.9883)[br]{$t$}
\FAProp(20.,5.)(16.,10.)(0.,){/Straight}{-1}
\FALabel(18.7303,7.98828)[bl]{$t$}
\FAProp(4.,10.)(10.,10.)(0.,){/Sine}{0}
\FALabel(7.,8.93)[t]{$\gamma, Z$}
\FAProp(16.,10.)(10.,10.)(0.,){/Sine}{0}
\FALabel(13.,8.93)[t]{$\gamma, Z$}
\FAProp(10.,10.)(10.,10.)(10.,15.5){/ScalarDash}{-1}
\FALabel(10.,16.57)[b]{$G$}
\FAVert(4.,10.){0}
\FAVert(16.,10.){0}
\FAVert(10.,10.){0}

\FADiagram{}
\FAProp(0.,15.)(4.,10.)(0.,){/Straight}{-1}
\FALabel(1.26965,12.0117)[tr]{$e$}
\FAProp(0.,5.)(4.,10.)(0.,){/Straight}{1}
\FALabel(2.73035,7.01172)[tl]{$e$}
\FAProp(20.,15.)(16.,10.)(0.,){/Straight}{1}
\FALabel(17.2697,12.9883)[br]{$t$}
\FAProp(20.,5.)(16.,10.)(0.,){/Straight}{-1}
\FALabel(18.7303,7.98828)[bl]{$t$}
\FAProp(4.,10.)(10.,10.)(0.,){/Sine}{0}
\FALabel(7.,8.93)[t]{$\gamma, Z$}
\FAProp(16.,10.)(10.,10.)(0.,){/Sine}{0}
\FALabel(13.,8.93)[t]{$\gamma, Z$}
\FAProp(10.,10.)(10.,10.)(10.,15.5){/Sine}{-1}
\FALabel(10.,16.57)[b]{$W$}
\FAVert(4.,10.){0}
\FAVert(16.,10.){0}
\FAVert(10.,10.){0}

\FADiagram{}
\FAProp(0.,15.)(3.,10.)(0.,){/Straight}{-1}
\FALabel(0.650886,12.1825)[tr]{$e$}
\FAProp(0.,5.)(3.,10.)(0.,){/Straight}{1}
\FALabel(2.34911,7.18253)[tl]{$e$}
\FAProp(20.,15.)(17.,10.)(0.,){/Straight}{1}
\FALabel(17.6509,12.8175)[br]{$t$}
\FAProp(20.,5.)(17.,10.)(0.,){/Straight}{-1}
\FALabel(19.3491,7.81747)[bl]{$t$}
\FAProp(3.,10.)(7.,10.)(0.,){/Sine}{0}
\FALabel(5.,11.07)[b]{$\gamma, Z$}
\FAProp(17.,10.)(13.,10.)(0.,){/Sine}{0}
\FALabel(15.,8.93)[t]{$\gamma, Z$}
\FAProp(7.,10.)(13.,10.)(0.8,){/Straight}{-1}
\FALabel(10.,6.53)[t]{$f_i$}
\FAProp(7.,10.)(13.,10.)(-0.8,){/Straight}{1}
\FALabel(10.,13.47)[b]{$f_i$}
\FAVert(3.,10.){0}
\FAVert(17.,10.){0}
\FAVert(7.,10.){0}
\FAVert(13.,10.){0}

\FADiagram{}
\FAProp(0.,15.)(3.,10.)(0.,){/Straight}{-1}
\FALabel(0.650886,12.1825)[tr]{$e$}
\FAProp(0.,5.)(3.,10.)(0.,){/Straight}{1}
\FALabel(2.34911,7.18253)[tl]{$e$}
\FAProp(20.,15.)(17.,10.)(0.,){/Straight}{1}
\FALabel(17.6509,12.8175)[br]{$t$}
\FAProp(20.,5.)(17.,10.)(0.,){/Straight}{-1}
\FALabel(19.3491,7.81747)[bl]{$t$}
\FAProp(3.,10.)(7.,10.)(0.,){/Sine}{0}
\FALabel(5.,11.07)[b]{$\gamma, Z$}
\FAProp(17.,10.)(13.,10.)(0.,){/Sine}{0}
\FALabel(15.,8.93)[t]{$\gamma, Z$}
\FAProp(7.,10.)(13.,10.)(0.8,){/GhostDash}{-1}
\FALabel(10.,6.53)[t]{$u_\pm$}
\FAProp(7.,10.)(13.,10.)(-0.8,){/GhostDash}{1}
\FALabel(10.,13.47)[b]{$u_\pm$}
\FAVert(3.,10.){0}
\FAVert(17.,10.){0}
\FAVert(7.,10.){0}
\FAVert(13.,10.){0}

\FADiagram{}
\FAProp(0.,15.)(3.,10.)(0.,){/Straight}{-1}
\FALabel(0.650886,12.1825)[tr]{$e$}
\FAProp(0.,5.)(3.,10.)(0.,){/Straight}{1}
\FALabel(2.34911,7.18253)[tl]{$e$}
\FAProp(20.,15.)(17.,10.)(0.,){/Straight}{1}
\FALabel(17.6509,12.8175)[br]{$t$}
\FAProp(20.,5.)(17.,10.)(0.,){/Straight}{-1}
\FALabel(19.3491,7.81747)[bl]{$t$}
\FAProp(3.,10.)(7.,10.)(0.,){/Sine}{0}
\FALabel(5.,11.07)[b]{$\gamma, Z$}
\FAProp(17.,10.)(13.,10.)(0.,){/Sine}{0}
\FALabel(15.,8.93)[t]{$\gamma, Z$}
\FAProp(7.,10.)(13.,10.)(0.8,){/ScalarDash}{-1}
\FALabel(10.,6.53)[t]{$G$}
\FAProp(7.,10.)(13.,10.)(-0.8,){/ScalarDash}{1}
\FALabel(10.,13.47)[b]{$G$}
\FAVert(3.,10.){0}
\FAVert(17.,10.){0}
\FAVert(7.,10.){0}
\FAVert(13.,10.){0}

\FADiagram{}
\FAProp(0.,15.)(3.,10.)(0.,){/Straight}{-1}
\FALabel(0.650886,12.1825)[tr]{$e$}
\FAProp(0.,5.)(3.,10.)(0.,){/Straight}{1}
\FALabel(2.34911,7.18253)[tl]{$e$}
\FAProp(20.,15.)(17.,10.)(0.,){/Straight}{1}
\FALabel(17.6509,12.8175)[br]{$t$}
\FAProp(20.,5.)(17.,10.)(0.,){/Straight}{-1}
\FALabel(19.3491,7.81747)[bl]{$t$}
\FAProp(3.,10.)(7.,10.)(0.,){/Sine}{0}
\FALabel(5.,11.07)[b]{$\gamma, Z$}
\FAProp(17.,10.)(13.,10.)(0.,){/Sine}{0}
\FALabel(15.,8.93)[t]{$\gamma, Z$}
\FAProp(7.,10.)(13.,10.)(0.8,){/ScalarDash}{-1}
\FALabel(10.,6.53)[t]{$G$}
\FAProp(7.,10.)(13.,10.)(-0.8,){/Sine}{1}
\FALabel(10.,13.47)[b]{$W$}
\FAVert(3.,10.){0}
\FAVert(17.,10.){0}
\FAVert(7.,10.){0}
\FAVert(13.,10.){0}

\FADiagram{}
\FAProp(0.,15.)(3.,10.)(0.,){/Straight}{-1}
\FALabel(0.650886,12.1825)[tr]{$e$}
\FAProp(0.,5.)(3.,10.)(0.,){/Straight}{1}
\FALabel(2.34911,7.18253)[tl]{$e$}
\FAProp(20.,15.)(17.,10.)(0.,){/Straight}{1}
\FALabel(17.6509,12.8175)[br]{$t$}
\FAProp(20.,5.)(17.,10.)(0.,){/Straight}{-1}
\FALabel(19.3491,7.81747)[bl]{$t$}
\FAProp(3.,10.)(7.,10.)(0.,){/Sine}{0}
\FALabel(5.,11.07)[b]{$\gamma, Z$}
\FAProp(17.,10.)(13.,10.)(0.,){/Sine}{0}
\FALabel(15.,8.93)[t]{$\gamma, Z$}
\FAProp(7.,10.)(13.,10.)(0.8,){/Sine}{-1}
\FALabel(10.,6.53)[t]{$W$}
\FAProp(7.,10.)(13.,10.)(-0.8,){/ScalarDash}{1}
\FALabel(10.,13.47)[b]{$G$}
\FAVert(3.,10.){0}
\FAVert(17.,10.){0}
\FAVert(7.,10.){0}
\FAVert(13.,10.){0}

\FADiagram{}
\FAProp(0.,15.)(3.,10.)(0.,){/Straight}{-1}
\FALabel(0.650886,12.1825)[tr]{$e$}
\FAProp(0.,5.)(3.,10.)(0.,){/Straight}{1}
\FALabel(2.34911,7.18253)[tl]{$e$}
\FAProp(20.,15.)(17.,10.)(0.,){/Straight}{1}
\FALabel(17.6509,12.8175)[br]{$t$}
\FAProp(20.,5.)(17.,10.)(0.,){/Straight}{-1}
\FALabel(19.3491,7.81747)[bl]{$t$}
\FAProp(3.,10.)(7.,10.)(0.,){/Sine}{0}
\FALabel(5.,11.07)[b]{$\gamma, Z$}
\FAProp(17.,10.)(13.,10.)(0.,){/Sine}{0}
\FALabel(15.,8.93)[t]{$\gamma, Z$}
\FAProp(7.,10.)(13.,10.)(0.8,){/Sine}{-1}
\FALabel(10.,6.53)[t]{$W$}
\FAProp(7.,10.)(13.,10.)(-0.8,){/Sine}{1}
\FALabel(10.,13.47)[b]{$W$}
\FAVert(3.,10.){0}
\FAVert(17.,10.){0}
\FAVert(7.,10.){0}
\FAVert(13.,10.){0}

\FADiagram{}
\FAProp(0.,15.)(4.,10.)(0.,){/Straight}{-1}
\FALabel(1.26965,12.0117)[tr]{$e$}
\FAProp(0.,5.)(4.,10.)(0.,){/Straight}{1}
\FALabel(2.73035,7.01172)[tl]{$e$}
\FAProp(20.,15.)(16.,10.)(0.,){/Straight}{1}
\FALabel(17.2697,12.9883)[br]{$t$}
\FAProp(20.,5.)(16.,10.)(0.,){/Straight}{-1}
\FALabel(18.7303,7.98828)[bl]{$t$}
\FAProp(4.,10.)(10.,10.)(0.,){/Sine}{0}
\FALabel(7.,8.93)[t]{$Z$}
\FAProp(16.,10.)(10.,10.)(0.,){/Sine}{0}
\FALabel(13.,8.93)[t]{$Z$}
\FAProp(10.,10.)(10.,10.)(10.,15.5){/ScalarDash}{0}
\FALabel(10.,16.32)[b]{$G^0, H$}
\FAVert(4.,10.){0}
\FAVert(16.,10.){0}
\FAVert(10.,10.){0}

\FADiagram{}
\FAProp(0.,15.)(3.,10.)(0.,){/Straight}{-1}
\FALabel(0.650886,12.1825)[tr]{$e$}
\FAProp(0.,5.)(3.,10.)(0.,){/Straight}{1}
\FALabel(2.34911,7.18253)[tl]{$e$}
\FAProp(20.,15.)(17.,10.)(0.,){/Straight}{1}
\FALabel(17.6509,12.8175)[br]{$t$}
\FAProp(20.,5.)(17.,10.)(0.,){/Straight}{-1}
\FALabel(19.3491,7.81747)[bl]{$t$}
\FAProp(3.,10.)(7.,10.)(0.,){/Sine}{0}
\FALabel(5.,11.07)[b]{$Z$}
\FAProp(17.,10.)(13.,10.)(0.,){/Sine}{0}
\FALabel(15.,8.93)[t]{$Z$}
\FAProp(7.,10.)(13.,10.)(0.8,){/Straight}{-1}
\FALabel(10.,6.53)[t]{$\nu_i$}
\FAProp(7.,10.)(13.,10.)(-0.8,){/Straight}{1}
\FALabel(10.,13.47)[b]{$\nu_i$}
\FAVert(3.,10.){0}
\FAVert(17.,10.){0}
\FAVert(7.,10.){0}
\FAVert(13.,10.){0}

\FADiagram{}
\FAProp(0.,15.)(3.,10.)(0.,){/Straight}{-1}
\FALabel(0.650886,12.1825)[tr]{$e$}
\FAProp(0.,5.)(3.,10.)(0.,){/Straight}{1}
\FALabel(2.34911,7.18253)[tl]{$e$}
\FAProp(20.,15.)(17.,10.)(0.,){/Straight}{1}
\FALabel(17.6509,12.8175)[br]{$t$}
\FAProp(20.,5.)(17.,10.)(0.,){/Straight}{-1}
\FALabel(19.3491,7.81747)[bl]{$t$}
\FAProp(3.,10.)(7.,10.)(0.,){/Sine}{0}
\FALabel(5.,11.07)[b]{$Z$}
\FAProp(17.,10.)(13.,10.)(0.,){/Sine}{0}
\FALabel(15.,8.93)[t]{$Z$}
\FAProp(7.,10.)(13.,10.)(0.8,){/ScalarDash}{0}
\FALabel(10.,6.78)[t]{$H$}
\FAProp(7.,10.)(13.,10.)(-0.8,){/ScalarDash}{0}
\FALabel(10.,13.22)[b]{$G^0$}
\FAVert(3.,10.){0}
\FAVert(17.,10.){0}
\FAVert(7.,10.){0}
\FAVert(13.,10.){0}

\FADiagram{}
\FAProp(0.,15.)(3.,10.)(0.,){/Straight}{-1}
\FALabel(0.650886,12.1825)[tr]{$e$}
\FAProp(0.,5.)(3.,10.)(0.,){/Straight}{1}
\FALabel(2.34911,7.18253)[tl]{$e$}
\FAProp(20.,15.)(17.,10.)(0.,){/Straight}{1}
\FALabel(17.6509,12.8175)[br]{$t$}
\FAProp(20.,5.)(17.,10.)(0.,){/Straight}{-1}
\FALabel(19.3491,7.81747)[bl]{$t$}
\FAProp(3.,10.)(7.,10.)(0.,){/Sine}{0}
\FALabel(5.,11.07)[b]{$Z$}
\FAProp(17.,10.)(13.,10.)(0.,){/Sine}{0}
\FALabel(15.,8.93)[t]{$Z$}
\FAProp(7.,10.)(13.,10.)(0.8,){/ScalarDash}{0}
\FALabel(10.,6.78)[t]{$H$}
\FAProp(7.,10.)(13.,10.)(-0.8,){/Sine}{0}
\FALabel(10.,13.47)[b]{$Z$}
\FAVert(3.,10.){0}
\FAVert(17.,10.){0}
\FAVert(7.,10.){0}
\FAVert(13.,10.){0}
\end{feynartspicture}
\end{scriptsize}

%% file: vert.tex
\unitlength=1bp%

\begin{scriptsize}
\begin{feynartspicture}(432,404)(4,4)
\FADiagram{}
\FAProp(0.,15.)(4.,10.)(0.,){/Straight}{-1}
\FALabel(1.26965,12.0117)[tr]{$e$}
\FAProp(0.,5.)(4.,10.)(0.,){/Straight}{1}
\FALabel(2.73035,7.01172)[tl]{$e$}
\FAProp(20.,15.)(16.,14.)(0.,){/Straight}{1}
\FALabel(17.6241,15.5237)[b]{$t$}
\FAProp(20.,5.)(16.,6.)(0.,){/Straight}{-1}
\FALabel(17.6241,4.47628)[t]{$t$}
\FAProp(4.,10.)(9.5,10.)(0.,){/Sine}{0}
\FALabel(6.75,11.07)[b]{$\gamma, Z$}
\FAProp(16.,14.)(16.,6.)(0.,){/ScalarDash}{0}
\FALabel(16.82,10.)[l]{\vtop{\hbox{$G^0,$}\hbox{$H$}}}
\FAProp(16.,14.)(9.5,10.)(0.,){/Straight}{1}
\FALabel(12.4176,12.8401)[br]{$t$}
\FAProp(16.,6.)(9.5,10.)(0.,){/Straight}{-1}
\FALabel(12.4176,7.15993)[tr]{$t$}
\FAVert(4.,10.){0}
\FAVert(16.,14.){0}
\FAVert(16.,6.){0}
\FAVert(9.5,10.){0}

\FADiagram{}
\FAProp(0.,15.)(4.,10.)(0.,){/Straight}{-1}
\FALabel(1.26965,12.0117)[tr]{$e$}
\FAProp(0.,5.)(4.,10.)(0.,){/Straight}{1}
\FALabel(2.73035,7.01172)[tl]{$e$}
\FAProp(20.,15.)(16.,14.)(0.,){/Straight}{1}
\FALabel(17.6241,15.5237)[b]{$t$}
\FAProp(20.,5.)(16.,6.)(0.,){/Straight}{-1}
\FALabel(17.6241,4.47628)[t]{$t$}
\FAProp(4.,10.)(9.5,10.)(0.,){/Sine}{0}
\FALabel(6.75,11.07)[b]{$\gamma, Z$}
\FAProp(16.,14.)(16.,6.)(0.,){/Sine}{0}
\FALabel(17.07,10.)[l]{\vtop{\hbox{$\gamma,$}\hbox{$Z$}}}
\FAProp(16.,14.)(9.5,10.)(0.,){/Straight}{1}
\FALabel(12.4176,12.8401)[br]{$t$}
\FAProp(16.,6.)(9.5,10.)(0.,){/Straight}{-1}
\FALabel(12.4176,7.15993)[tr]{$t$}
\FAVert(4.,10.){0}
\FAVert(16.,14.){0}
\FAVert(16.,6.){0}
\FAVert(9.5,10.){0}

\FADiagram{}
\FAProp(0.,15.)(4.,10.)(0.,){/Straight}{-1}
\FALabel(1.26965,12.0117)[tr]{$e$}
\FAProp(0.,5.)(4.,10.)(0.,){/Straight}{1}
\FALabel(2.73035,7.01172)[tl]{$e$}
\FAProp(20.,15.)(16.,14.)(0.,){/Straight}{1}
\FALabel(17.6241,15.5237)[b]{$t$}
\FAProp(20.,5.)(16.,6.)(0.,){/Straight}{-1}
\FALabel(17.6241,4.47628)[t]{$t$}
\FAProp(4.,10.)(9.5,10.)(0.,){/Sine}{0}
\FALabel(6.75,11.07)[b]{$\gamma, Z$}
\FAProp(16.,14.)(16.,6.)(0.,){/ScalarDash}{-1}
\FALabel(17.07,10.)[l]{$G$}
\FAProp(16.,14.)(9.5,10.)(0.,){/Straight}{1}
\FALabel(12.4176,12.8401)[br]{$b$}
\FAProp(16.,6.)(9.5,10.)(0.,){/Straight}{-1}
\FALabel(12.4176,7.15993)[tr]{$b$}
\FAVert(4.,10.){0}
\FAVert(16.,14.){0}
\FAVert(16.,6.){0}
\FAVert(9.5,10.){0}

\FADiagram{}
\FAProp(0.,15.)(4.,10.)(0.,){/Straight}{-1}
\FALabel(1.26965,12.0117)[tr]{$e$}
\FAProp(0.,5.)(4.,10.)(0.,){/Straight}{1}
\FALabel(2.73035,7.01172)[tl]{$e$}
\FAProp(20.,15.)(16.,14.)(0.,){/Straight}{1}
\FALabel(17.6241,15.5237)[b]{$t$}
\FAProp(20.,5.)(16.,6.)(0.,){/Straight}{-1}
\FALabel(17.6241,4.47628)[t]{$t$}
\FAProp(4.,10.)(9.5,10.)(0.,){/Sine}{0}
\FALabel(6.75,11.07)[b]{$\gamma, Z$}
\FAProp(16.,14.)(16.,6.)(0.,){/Sine}{-1}
\FALabel(17.07,10.)[l]{$W$}
\FAProp(16.,14.)(9.5,10.)(0.,){/Straight}{1}
\FALabel(12.4176,12.8401)[br]{$b$}
\FAProp(16.,6.)(9.5,10.)(0.,){/Straight}{-1}
\FALabel(12.4176,7.15993)[tr]{$b$}
\FAVert(4.,10.){0}
\FAVert(16.,14.){0}
\FAVert(16.,6.){0}
\FAVert(9.5,10.){0}

\FADiagram{}
\FAProp(0.,15.)(4.,10.)(0.,){/Straight}{-1}
\FALabel(1.26965,12.0117)[tr]{$e$}
\FAProp(0.,5.)(4.,10.)(0.,){/Straight}{1}
\FALabel(2.73035,7.01172)[tl]{$e$}
\FAProp(20.,15.)(16.,14.)(0.,){/Straight}{1}
\FALabel(17.6241,15.5237)[b]{$t$}
\FAProp(20.,5.)(16.,6.)(0.,){/Straight}{-1}
\FALabel(17.6241,4.47628)[t]{$t$}
\FAProp(4.,10.)(9.5,10.)(0.,){/Sine}{0}
\FALabel(6.75,11.07)[b]{$\gamma, Z$}
\FAProp(16.,14.)(16.,6.)(0.,){/Straight}{1}
\FALabel(17.07,10.)[l]{$b$}
\FAProp(16.,14.)(9.5,10.)(0.,){/ScalarDash}{-1}
\FALabel(12.4176,12.8401)[br]{$G$}
\FAProp(16.,6.)(9.5,10.)(0.,){/ScalarDash}{1}
\FALabel(12.4176,7.15993)[tr]{$G$}
\FAVert(4.,10.){0}
\FAVert(16.,14.){0}
\FAVert(16.,6.){0}
\FAVert(9.5,10.){0}

\FADiagram{}
\FAProp(0.,15.)(4.,10.)(0.,){/Straight}{-1}
\FALabel(1.26965,12.0117)[tr]{$e$}
\FAProp(0.,5.)(4.,10.)(0.,){/Straight}{1}
\FALabel(2.73035,7.01172)[tl]{$e$}
\FAProp(20.,15.)(16.,14.)(0.,){/Straight}{1}
\FALabel(17.6241,15.5237)[b]{$t$}
\FAProp(20.,5.)(16.,6.)(0.,){/Straight}{-1}
\FALabel(17.6241,4.47628)[t]{$t$}
\FAProp(4.,10.)(9.5,10.)(0.,){/Sine}{0}
\FALabel(6.75,11.07)[b]{$\gamma, Z$}
\FAProp(16.,14.)(16.,6.)(0.,){/Straight}{1}
\FALabel(17.07,10.)[l]{$b$}
\FAProp(16.,14.)(9.5,10.)(0.,){/ScalarDash}{-1}
\FALabel(12.4176,12.8401)[br]{$G$}
\FAProp(16.,6.)(9.5,10.)(0.,){/Sine}{1}
\FALabel(12.4176,7.15993)[tr]{$W$}
\FAVert(4.,10.){0}
\FAVert(16.,14.){0}
\FAVert(16.,6.){0}
\FAVert(9.5,10.){0}

\FADiagram{}
\FAProp(0.,15.)(4.,10.)(0.,){/Straight}{-1}
\FALabel(1.26965,12.0117)[tr]{$e$}
\FAProp(0.,5.)(4.,10.)(0.,){/Straight}{1}
\FALabel(2.73035,7.01172)[tl]{$e$}
\FAProp(20.,15.)(16.,14.)(0.,){/Straight}{1}
\FALabel(17.6241,15.5237)[b]{$t$}
\FAProp(20.,5.)(16.,6.)(0.,){/Straight}{-1}
\FALabel(17.6241,4.47628)[t]{$t$}
\FAProp(4.,10.)(9.5,10.)(0.,){/Sine}{0}
\FALabel(6.75,11.07)[b]{$\gamma, Z$}
\FAProp(16.,14.)(16.,6.)(0.,){/Straight}{1}
\FALabel(17.07,10.)[l]{$b$}
\FAProp(16.,14.)(9.5,10.)(0.,){/Sine}{-1}
\FALabel(12.4176,12.8401)[br]{$W$}
\FAProp(16.,6.)(9.5,10.)(0.,){/ScalarDash}{1}
\FALabel(12.4176,7.15993)[tr]{$G$}
\FAVert(4.,10.){0}
\FAVert(16.,14.){0}
\FAVert(16.,6.){0}
\FAVert(9.5,10.){0}

\FADiagram{}
\FAProp(0.,15.)(4.,10.)(0.,){/Straight}{-1}
\FALabel(1.26965,12.0117)[tr]{$e$}
\FAProp(0.,5.)(4.,10.)(0.,){/Straight}{1}
\FALabel(2.73035,7.01172)[tl]{$e$}
\FAProp(20.,15.)(16.,14.)(0.,){/Straight}{1}
\FALabel(17.6241,15.5237)[b]{$t$}
\FAProp(20.,5.)(16.,6.)(0.,){/Straight}{-1}
\FALabel(17.6241,4.47628)[t]{$t$}
\FAProp(4.,10.)(9.5,10.)(0.,){/Sine}{0}
\FALabel(6.75,11.07)[b]{$\gamma, Z$}
\FAProp(16.,14.)(16.,6.)(0.,){/Straight}{1}
\FALabel(17.07,10.)[l]{$b$}
\FAProp(16.,14.)(9.5,10.)(0.,){/Sine}{-1}
\FALabel(12.4176,12.8401)[br]{$W$}
\FAProp(16.,6.)(9.5,10.)(0.,){/Sine}{1}
\FALabel(12.4176,7.15993)[tr]{$W$}
\FAVert(4.,10.){0}
\FAVert(16.,14.){0}
\FAVert(16.,6.){0}
\FAVert(9.5,10.){0}

\FADiagram{}
\FAProp(0.,15.)(4.,10.)(0.,){/Straight}{-1}
\FALabel(1.26965,12.0117)[tr]{$e$}
\FAProp(0.,5.)(4.,10.)(0.,){/Straight}{1}
\FALabel(2.73035,7.01172)[tl]{$e$}
\FAProp(20.,15.)(16.,14.)(0.,){/Straight}{1}
\FALabel(17.6241,15.5237)[b]{$t$}
\FAProp(20.,5.)(16.,6.)(0.,){/Straight}{-1}
\FALabel(17.6241,4.47628)[t]{$t$}
\FAProp(4.,10.)(9.5,10.)(0.,){/Sine}{0}
\FALabel(6.75,11.07)[b]{$Z$}
\FAProp(16.,14.)(16.,6.)(0.,){/Straight}{1}
\FALabel(17.07,10.)[l]{$t$}
\FAProp(16.,14.)(9.5,10.)(0.,){/ScalarDash}{0}
\FALabel(12.5487,12.6272)[br]{$H$}
\FAProp(16.,6.)(9.5,10.)(0.,){/ScalarDash}{0}
\FALabel(12.5487,7.37284)[tr]{$G^0$}
\FAVert(4.,10.){0}
\FAVert(16.,14.){0}
\FAVert(16.,6.){0}
\FAVert(9.5,10.){0}

\FADiagram{}
\FAProp(0.,15.)(4.,10.)(0.,){/Straight}{-1}
\FALabel(1.26965,12.0117)[tr]{$e$}
\FAProp(0.,5.)(4.,10.)(0.,){/Straight}{1}
\FALabel(2.73035,7.01172)[tl]{$e$}
\FAProp(20.,15.)(16.,14.)(0.,){/Straight}{1}
\FALabel(17.6241,15.5237)[b]{$t$}
\FAProp(20.,5.)(16.,6.)(0.,){/Straight}{-1}
\FALabel(17.6241,4.47628)[t]{$t$}
\FAProp(4.,10.)(9.5,10.)(0.,){/Sine}{0}
\FALabel(6.75,11.07)[b]{$Z$}
\FAProp(16.,14.)(16.,6.)(0.,){/Straight}{1}
\FALabel(17.07,10.)[l]{$t$}
\FAProp(16.,14.)(9.5,10.)(0.,){/ScalarDash}{0}
\FALabel(12.5487,12.6272)[br]{$H$}
\FAProp(16.,6.)(9.5,10.)(0.,){/Sine}{0}
\FALabel(12.4176,7.15993)[tr]{$Z$}
\FAVert(4.,10.){0}
\FAVert(16.,14.){0}
\FAVert(16.,6.){0}
\FAVert(9.5,10.){0}

\FADiagram{}
\FAProp(0.,15.)(4.,10.)(0.,){/Straight}{-1}
\FALabel(1.26965,12.0117)[tr]{$e$}
\FAProp(0.,5.)(4.,10.)(0.,){/Straight}{1}
\FALabel(2.73035,7.01172)[tl]{$e$}
\FAProp(20.,15.)(16.,14.)(0.,){/Straight}{1}
\FALabel(17.6241,15.5237)[b]{$t$}
\FAProp(20.,5.)(16.,6.)(0.,){/Straight}{-1}
\FALabel(17.6241,4.47628)[t]{$t$}
\FAProp(4.,10.)(9.5,10.)(0.,){/Sine}{0}
\FALabel(6.75,11.07)[b]{$Z$}
\FAProp(16.,14.)(16.,6.)(0.,){/Straight}{1}
\FALabel(17.07,10.)[l]{$t$}
\FAProp(16.,14.)(9.5,10.)(0.,){/ScalarDash}{0}
\FALabel(12.5487,12.6272)[br]{$G^0$}
\FAProp(16.,6.)(9.5,10.)(0.,){/ScalarDash}{0}
\FALabel(12.5487,7.37284)[tr]{$H$}
\FAVert(4.,10.){0}
\FAVert(16.,14.){0}
\FAVert(16.,6.){0}
\FAVert(9.5,10.){0}

\FADiagram{}
\FAProp(0.,15.)(4.,10.)(0.,){/Straight}{-1}
\FALabel(1.26965,12.0117)[tr]{$e$}
\FAProp(0.,5.)(4.,10.)(0.,){/Straight}{1}
\FALabel(2.73035,7.01172)[tl]{$e$}
\FAProp(20.,15.)(16.,14.)(0.,){/Straight}{1}
\FALabel(17.6241,15.5237)[b]{$t$}
\FAProp(20.,5.)(16.,6.)(0.,){/Straight}{-1}
\FALabel(17.6241,4.47628)[t]{$t$}
\FAProp(4.,10.)(9.5,10.)(0.,){/Sine}{0}
\FALabel(6.75,11.07)[b]{$Z$}
\FAProp(16.,14.)(16.,6.)(0.,){/Straight}{1}
\FALabel(17.07,10.)[l]{$t$}
\FAProp(16.,14.)(9.5,10.)(0.,){/Sine}{0}
\FALabel(12.4176,12.8401)[br]{$Z$}
\FAProp(16.,6.)(9.5,10.)(0.,){/ScalarDash}{0}
\FALabel(12.5487,7.37284)[tr]{$H$}
\FAVert(4.,10.){0}
\FAVert(16.,14.){0}
\FAVert(16.,6.){0}
\FAVert(9.5,10.){0}

\FADiagram{}
\FAProp(0.,15.)(4.,14.)(0.,){/Straight}{-1}
\FALabel(2.37593,15.5237)[b]{$e$}
\FAProp(0.,5.)(4.,6.)(0.,){/Straight}{1}
\FALabel(2.37593,4.47628)[t]{$e$}
\FAProp(20.,15.)(16.,10.)(0.,){/Straight}{1}
\FALabel(17.2697,12.9883)[br]{$t$}
\FAProp(20.,5.)(16.,10.)(0.,){/Straight}{-1}
\FALabel(18.7303,7.98828)[bl]{$t$}
\FAProp(16.,10.)(10.5,10.)(0.,){/Sine}{0}
\FALabel(13.25,8.93)[t]{$\gamma, Z$}
\FAProp(4.,14.)(4.,6.)(0.,){/Sine}{0}
\FALabel(2.93,10.)[r]{$\gamma, Z$}
\FAProp(4.,14.)(10.5,10.)(0.,){/Straight}{-1}
\FALabel(7.58235,12.8401)[bl]{$e$}
\FAProp(4.,6.)(10.5,10.)(0.,){/Straight}{1}
\FALabel(7.58235,7.15993)[tl]{$e$}
\FAVert(4.,14.){0}
\FAVert(4.,6.){0}
\FAVert(16.,10.){0}
\FAVert(10.5,10.){0}

\FADiagram{}
\FAProp(0.,15.)(4.,14.)(0.,){/Straight}{-1}
\FALabel(2.37593,15.5237)[b]{$e$}
\FAProp(0.,5.)(4.,6.)(0.,){/Straight}{1}
\FALabel(2.37593,4.47628)[t]{$e$}
\FAProp(20.,15.)(16.,10.)(0.,){/Straight}{1}
\FALabel(17.2697,12.9883)[br]{$t$}
\FAProp(20.,5.)(16.,10.)(0.,){/Straight}{-1}
\FALabel(18.7303,7.98828)[bl]{$t$}
\FAProp(16.,10.)(10.5,10.)(0.,){/Sine}{0}
\FALabel(13.25,8.93)[t]{$\gamma, Z$}
\FAProp(4.,14.)(4.,6.)(0.,){/Straight}{-1}
\FALabel(2.93,10.)[r]{$\nu_e$}
\FAProp(4.,14.)(10.5,10.)(0.,){/Sine}{-1}
\FALabel(7.58235,12.8401)[bl]{$W$}
\FAProp(4.,6.)(10.5,10.)(0.,){/Sine}{1}
\FALabel(7.58235,7.15993)[tl]{$W$}
\FAVert(4.,14.){0}
\FAVert(4.,6.){0}
\FAVert(16.,10.){0}
\FAVert(10.5,10.){0}

\FADiagram{}
\FAProp(0.,15.)(4.,14.)(0.,){/Straight}{-1}
\FALabel(2.37593,15.5237)[b]{$e$}
\FAProp(0.,5.)(4.,6.)(0.,){/Straight}{1}
\FALabel(2.37593,4.47628)[t]{$e$}
\FAProp(20.,15.)(16.,10.)(0.,){/Straight}{1}
\FALabel(17.2697,12.9883)[br]{$t$}
\FAProp(20.,5.)(16.,10.)(0.,){/Straight}{-1}
\FALabel(18.7303,7.98828)[bl]{$t$}
\FAProp(16.,10.)(10.5,10.)(0.,){/Sine}{0}
\FALabel(13.25,8.93)[t]{$Z$}
\FAProp(4.,14.)(4.,6.)(0.,){/Sine}{-1}
\FALabel(2.93,10.)[r]{$W$}
\FAProp(4.,14.)(10.5,10.)(0.,){/Straight}{-1}
\FALabel(7.58235,12.8401)[bl]{$\nu_e$}
\FAProp(4.,6.)(10.5,10.)(0.,){/Straight}{1}
\FALabel(7.58235,7.15993)[tl]{$\nu_e$}
\FAVert(4.,14.){0}
\FAVert(4.,6.){0}
\FAVert(16.,10.){0}
\FAVert(10.5,10.){0}
\end{feynartspicture}
\end{scriptsize}

%% file: real.tex
\unitlength=1bp%

\begin{scriptsize}
\begin{feynartspicture}(432,101)(4,1)
\FADiagram{}
\FAProp(0.,15.)(5.,10.)(0.,){/Straight}{-1}
\FALabel(1.88398,11.884)[tr]{$e$}
\FAProp(0.,5.)(5.,10.)(0.,){/Straight}{1}
\FALabel(3.11602,6.88398)[tl]{$e$}
\FAProp(20.,15.)(17.,12.5)(0.,){/Straight}{1}
\FALabel(17.9877,14.4607)[br]{$t$}
\FAProp(20.,5.)(14.,10.)(0.,){/Straight}{-1}
\FALabel(16.4877,6.78926)[tr]{$t$}
\FAProp(20.,10.)(17.,12.5)(0.,){/Sine}{0}
\FALabel(17.9877,10.5393)[tr]{$\gamma$}
\FAProp(5.,10.)(14.,10.)(0.,){/Sine}{0}
\FALabel(9.5,8.93)[t]{$\gamma, Z$}
\FAProp(17.,12.5)(14.,10.)(0.,){/Straight}{1}
\FALabel(14.9877,11.9607)[br]{$t$}
\FAVert(5.,10.){0}
\FAVert(17.,12.5){0}
\FAVert(14.,10.){0}

\FADiagram{}
\FAProp(0.,15.)(4.,10.)(0.,){/Straight}{-1}
\FALabel(1.26965,12.0117)[tr]{$e$}
\FAProp(0.,5.)(4.,10.)(0.,){/Straight}{1}
\FALabel(2.73035,7.01172)[tl]{$e$}
\FAProp(20.,15.)(14.,10.)(0.,){/Straight}{1}
\FALabel(16.4877,13.2107)[br]{$t$}
\FAProp(20.,5.)(17.,7.5)(0.,){/Straight}{-1}
\FALabel(17.9877,5.53926)[tr]{$t$}
\FAProp(20.,10.)(17.,7.5)(0.,){/Sine}{0}
\FALabel(17.9877,9.46074)[br]{$\gamma$}
\FAProp(4.,10.)(14.,10.)(0.,){/Sine}{0}
\FALabel(9.,11.07)[b]{$\gamma, Z$}
\FAProp(14.,10.)(17.,7.5)(0.,){/Straight}{1}
\FALabel(14.9877,8.03926)[tr]{$t$}
\FAVert(4.,10.){0}
\FAVert(14.,10.){0}
\FAVert(17.,7.5){0}

\FADiagram{}
\FAProp(0.,15.)(3.,12.5)(0.,){/Straight}{-1}
\FALabel(0.987714,13.0393)[tr]{$e$}
\FAProp(0.,5.)(6.,10.)(0.,){/Straight}{1}
\FALabel(3.51229,6.78926)[tl]{$e$}
\FAProp(20.,15.)(15.,10.)(0.,){/Straight}{1}
\FALabel(16.884,13.116)[br]{$t$}
\FAProp(20.,5.)(15.,10.)(0.,){/Straight}{-1}
\FALabel(18.116,8.11602)[bl]{$t$}
\FAProp(6.,15.)(3.,12.5)(0.,){/Sine}{0}
\FALabel(3.98771,14.4607)[br]{$\gamma$}
\FAProp(3.,12.5)(6.,10.)(0.,){/Straight}{-1}
\FALabel(3.98771,10.5393)[tr]{$e$}
\FAProp(6.,10.)(15.,10.)(0.,){/Sine}{0}
\FALabel(10.5,8.93)[t]{$\gamma, Z$}
\FAVert(3.,12.5){0}
\FAVert(6.,10.){0}
\FAVert(15.,10.){0}

\FADiagram{}
\FAProp(0.,15.)(6.,10.)(0.,){/Straight}{-1}
\FALabel(3.51229,13.2107)[bl]{$e$}
\FAProp(0.,5.)(3.,7.5)(0.,){/Straight}{1}
\FALabel(0.987714,6.96074)[br]{$e$}
\FAProp(20.,15.)(15.,10.)(0.,){/Straight}{1}
\FALabel(16.884,13.116)[br]{$t$}
\FAProp(20.,5.)(15.,10.)(0.,){/Straight}{-1}
\FALabel(18.116,8.11602)[bl]{$t$}
\FAProp(6.,5.)(3.,7.5)(0.,){/Sine}{0}
\FALabel(5.01229,6.96074)[bl]{$\gamma$}
\FAProp(6.,10.)(3.,7.5)(0.,){/Straight}{-1}
\FALabel(3.98771,9.46074)[br]{$e$}
\FAProp(6.,10.)(15.,10.)(0.,){/Sine}{0}
\FALabel(10.5,11.07)[b]{$\gamma, Z$}
\FAVert(6.,10.){0}
\FAVert(3.,7.5){0}
\FAVert(15.,10.){0}
\end{feynartspicture}
\end{scriptsize}

%% file: box.tex
\unitlength=1bp%

\begin{scriptsize}
\begin{feynartspicture}(432,101)(4,1)
\FADiagram{}
\FAProp(0.,15.)(5.5,14.5)(0.,){/Straight}{-1}
\FALabel(2.89033,15.8136)[b]{$e$}
\FAProp(0.,5.)(5.5,5.5)(0.,){/Straight}{1}
\FALabel(2.89033,4.18637)[t]{$e$}
\FAProp(20.,15.)(14.5,14.5)(0.,){/Straight}{1}
\FALabel(17.1097,15.8136)[b]{$t$}
\FAProp(20.,5.)(14.5,5.5)(0.,){/Straight}{-1}
\FALabel(17.1097,4.18637)[t]{$t$}
\FAProp(5.5,14.5)(5.5,5.5)(0.,){/Straight}{-1}
\FALabel(4.43,10.)[r]{$e$}
\FAProp(5.5,14.5)(14.5,14.5)(0.,){/Sine}{0}
\FALabel(10.,15.57)[b]{$\gamma, Z$}
\FAProp(5.5,5.5)(14.5,5.5)(0.,){/Sine}{0}
\FALabel(10.,4.43)[t]{$\gamma, Z$}
\FAProp(14.5,14.5)(14.5,5.5)(0.,){/Straight}{1}
\FALabel(15.57,10.)[l]{$t$}
\FAVert(5.5,14.5){0}
\FAVert(5.5,5.5){0}
\FAVert(14.5,14.5){0}
\FAVert(14.5,5.5){0}

\FADiagram{}
\FAProp(0.,15.)(5.5,14.5)(0.,){/Straight}{-1}
\FALabel(2.89033,15.8136)[b]{$e$}
\FAProp(0.,5.)(5.5,5.5)(0.,){/Straight}{1}
\FALabel(2.89033,4.18637)[t]{$e$}
\FAProp(20.,15.)(14.5,5.5)(0.,){/Straight}{1}
\FALabel(18.2636,13.1453)[br]{$t$}
\FAProp(20.,5.)(14.5,14.5)(0.,){/Straight}{-1}
\FALabel(18.7364,8.14526)[bl]{$t$}
\FAProp(5.5,14.5)(5.5,5.5)(0.,){/Straight}{-1}
\FALabel(4.43,10.)[r]{$e$}
\FAProp(5.5,14.5)(14.5,14.5)(0.,){/Sine}{0}
\FALabel(10.,15.57)[b]{$\gamma, Z$}
\FAProp(5.5,5.5)(14.5,5.5)(0.,){/Sine}{0}
\FALabel(10.,4.43)[t]{$\gamma, Z$}
\FAProp(14.5,5.5)(14.5,14.5)(0.,){/Straight}{1}
\FALabel(13.43,10.)[r]{$t$}
\FAVert(5.5,14.5){0}
\FAVert(5.5,5.5){0}
\FAVert(14.5,5.5){0}
\FAVert(14.5,14.5){0}

\FADiagram{}
\FAProp(0.,15.)(5.5,14.5)(0.,){/Straight}{-1}
\FALabel(2.89033,15.8136)[b]{$e$}
\FAProp(0.,5.)(5.5,5.5)(0.,){/Straight}{1}
\FALabel(2.89033,4.18637)[t]{$e$}
\FAProp(20.,15.)(14.5,5.5)(0.,){/Straight}{1}
\FALabel(18.2636,13.1453)[br]{$t$}
\FAProp(20.,5.)(14.5,14.5)(0.,){/Straight}{-1}
\FALabel(18.7364,8.14526)[bl]{$t$}
\FAProp(5.5,14.5)(5.5,5.5)(0.,){/Straight}{-1}
\FALabel(4.43,10.)[r]{$\nu_e$}
\FAProp(5.5,14.5)(14.5,14.5)(0.,){/Sine}{-1}
\FALabel(10.,15.57)[b]{$W$}
\FAProp(5.5,5.5)(14.5,5.5)(0.,){/Sine}{1}
\FALabel(10.,4.43)[t]{$W$}
\FAProp(14.5,5.5)(14.5,14.5)(0.,){/Straight}{1}
\FALabel(13.43,10.)[r]{$b$}
\FAVert(5.5,14.5){0}
\FAVert(5.5,5.5){0}
\FAVert(14.5,5.5){0}
\FAVert(14.5,14.5){0}
\end{feynartspicture}
\end{scriptsize}